# THE EARTH's ABSOLUTE GRAVITATION POTENTIAL FUNCTION IN THE PROSPECT «GRAVITATIONAL POTENTIAL METERING» OF GEOLOGICAL OBJECTS AND EARTHQUAKE CENTERS

Aleksandr Fridrikson and Marina Kasatochkina

Abstract:  The direct problem of the detection of the Earth's absolute gravitation potential maximum value  (MGP) was solved. The inverse problem finding of the Earth maximum gravitation  (where there is a maximum of gravitation field intensity  and a potential function has a «bending point») with the help of MGP was solved as well.
The obtained results show that the revealed Earth maximum gravitation coincides quite strictly with the cseismic D" layer on the border of the inner and outer (liquid) core.
The validity of the method of an absolute gravitation potential detection by the equal-potential velocity was proved as  «gravitation potential measurement» or «Vs-gravity method».
The prospects of this method for detecting of low-power  or distant geological objects with abnormal density and the possible earthquakes with low density  was shown.

---------------------------------------------------------------------------------------------------------

# ФУНКЦИЯ АБСОЛЮТНОГО ГРАВИТАЦИОННОГО ПОТЕНЦИАЛА ЗЕМЛИ В ПЕРСПЕКТИВЕ «ГРАВИПОТЕНЦИАЛОМЕТРИИ» ГЕОЛОГИЧЕСКИХ ОБЪЕКТОВ И ОЧАГОВ ЗЕМЛЕТРЯСЕНИЙ

Александр Фридрихсон и Марина Касаточкина

Резюме   Решена прямая задача по нахождению максимального значения абсолютного гравитационного потенциала (МГП) Земли. Решена обратная задача по нахождению через МГП  - гравитационного максимума Земли, где напряжённость поля тяготения достигает максимального значения, а потенциальная функция имеет точку перегиба.
Полученные результаты показывают, что выявленный  гравитационный максимум Земли весьма точно  совпадает с сейсмологическим слоем D" - на границе внешнего (жидкого) и внутреннего ядра.
Подтверждена работоспособность метода определения абсолютного гравитационного потенциала через значение эквипотенциальной скорости в точке определения, названного «гравипотенциалометрия» или «$v_S$-гравиметрия». Показана перспектива данного метода в выявления маломощных или удалённых геологических объектов аномальной плотности, а также мало плотных очагов возможных землетрясений.

***

Рассмотрим гравитационный потенциал поля тяготения однородного масс-источника $M$, радиуса $R$, с точкой определения в сферической координате $r$, внутри: $(0 \leq r \leq R)$ и вне : $(R < r < \infty)$ источника поля тяготения - как функцию трех переменных с частными производными, две из которых априори очевидны, а третья мало исследована и, как правило, не принимается в расчёт:

(0) $$U = U(M, r, R)$$

$$\frac{\partial U}{\partial M} > 0., \quad \frac{\partial U}{\partial r} > 0., \quad \frac{\partial U(r = const)}{\partial R} < 0$$

Договоримся, что речь идёт об абсолютном значении потенциала. Будем исходить из закона сохранения энергии примененного к пробной массе $m$ в изотропном гравитационном поле масс-источника $M_R$, ограниченного поверхностью $4\pi R^2$

(1) $$E_0 = \gamma m \int_0^R \frac{M_r}{r^2} dr + \gamma m M_R \int_R^\infty \frac{dr}{r^2} = \boldsymbol{const}$$

Где:
$\mathrm{M_r}$ - масса заключенная внутри поверхности $S_r = 4\pi r^2$ при $r \leq R$
$\mathrm{E_0}$ - минимальная энергия, необходимая для перемещения пробной массы
    m из центра масс-источника $\mathrm{M_R}$ в бесконечность.

Определим гравитационный потенциал на всем интервале $[0 < r < \infty]$ как минимальную удельную (на единицу массы) энергию, необходимую, чтобы переместить пробное тело из центра масс в заданную координату $r > R$. При этом, введём плотность масс-источника $\rho = \boldsymbol{const}$

(2) $$U(M, r, R) = \gamma \left(\frac{4}{3}\right) \pi \rho \int_0^R r dr + \gamma M_R \int_R^r \frac{1}{r^2} dr = \left[\left(\frac{2}{3}\right) \gamma \rho \pi R^2 - \frac{\gamma M_R}{r} + \frac{\gamma M_R}{R}\right]$$

Рассмотрим некоторые следствия полученного выражения. Прежде всего, надо выделить точку $r = R$. Здесь на стыке параболической и гиперболической частей, функция абсолютного потенциала априори имеет точку перегиба – то есть, максимальную первую производную и «нулевую» вторую производную.

Из (1) и (2) очевидно также, что кинетический потенциал, или удельная кинетическая энергия, которую приобретёт пробное тело, падая по нормали из бесконечности на некоторую поверхность $S_r$ - есть разность гравитационных потенциалов бесконечно удаленной точки и точки, куда «приземлится», под действием гравитации, единичная масса. Закон сохранения, при этом, имеет весьма простой вид:

(3) $$U_\infty = U_i + \frac{(\boldsymbol{v}_{\infty r})^2}{2} = \boldsymbol{const}$$

Где: $U_i$ и $(\boldsymbol{v}_{\infty r})^2/2 = K_i$ - гравитационный и кинетический потенциал.

Несложно видеть, что «нормально» падая из бесконечности в точку $R$, то есть на граничную поверхность $S_R$ масс-источника поля, пробное тело проходит значения:

$$(3-1) \qquad |\mathbf{v}|_{\infty r} \;=\; \sqrt{\frac{\gamma M_R}{r}} \;=\; |\mathbf{v}_{1k}|$$

«первых космических» скоростей приуроченных к круговым орбитам $2\pi r \in S_r$.
По определению первая космическая скорость достаточна для «нормального» удаления из заданной точки в бесконечность, где обращается в нуль. Из того же определения следует, что первая космическая скорость, направленная не по нормали, но тангенциально радиус-вектору $r$ - достаточна, чтобы пробное тело двигалось по стационарной круговой орбите сколь угодно долго. То есть бесконечно, но уже во времени:

$$(3-2) \qquad r(t) = \boldsymbol{const}., \quad \mathbf{v}_{1k} = \boldsymbol{v}_S(t) = \boldsymbol{const} \text{ при } t \to \infty.,$$

Где: $\boldsymbol{v}_S$ - скорость «бесконечного» нахождения на эквипотенциали $S_r$ или эквипотенциальная скорость.

Подставляя (3-2) в (3), имеем:

$$(4-1) \qquad U_i \;=\; U_\infty \;-\; \frac{\boldsymbol{v}_S{}^2}{2}$$

Где: $(\boldsymbol{v}_S{}^2/2) = C_i$ - кинетический потенциал необходимый для «бесконечного» пребывания на заданной эквипотенциальной поверхности гравитационного поля., или сокращенно – «эквипотенциал»

*Выводы:*
1) $\boldsymbol{v}_S(r,t)$ – представляет собой измеряемый баллистическими гравиметрами параметр гравитационного поля, аномалии и временные вариации которого выражают аномалии и вариации абсолютного гравитационного потенциала.
2) Аномалии плотности $\Delta\rho$ удаленных или маломощных $h \gg \Delta h$ объектов эффективнее отражены и могут быть выявлены по аномалии $(U_i - U_0)$ абсолютного гравитационного потенциала, нежели напряжённости гравитационного поля $(g_i - g_0)$. В этом несложно убедиться, выделив и сравнив коэффициенты линейных функций приращения потенциала и напряженности гравитационного поля от приращения плотности и приведённого (к размеру источника аномалии) радиус-вектора точки наблюдения:

$$(4-1-1) \qquad U_i - U_0 \;=\; \Delta U_i = k_1 \Delta \rho_i \,;\quad g_i - g_0 \;=\; \Delta g_i = k_2 \Delta \rho_i$$

$$k_1 = (2/3)\pi\gamma(r/r_0)^2 \;>\; (4/3)\pi\gamma(r/r_0) \;=\; k_2$$

$$\text{при всех} \quad r > 2r_0; \quad \text{или} \quad h > \Delta h$$

3) Значения эквипотенциальных скоростей могут быть использованы не только в гравиметрии, но и в определении поверхностей уровня, при решении геодезических и глобальных геодезических задач. Действительно, если $r$ – расстояние до центра тяготеющей массы, то в соответствии с известным уравнением небесной механики:
$g(r) = \mathbf{v}_{1k}{}^2/r$ получаем выражение и для решения данной задачи:

$$(4-1-2) \qquad r(\boldsymbol{v}_S) = \frac{{\boldsymbol{v}_{Sr}}^2}{g_r}$$

Таким образом, $\boldsymbol{v}_S$-гравиметрия при определенной технологической переработке баллистического способа, даст ощутимый результат в самых разных областях - особенно на удаленных и маломощных плотностных аномалиях. Прежде всего, таких, как очаги землетрясений, нефтегазоносные пласты, глубокозалегающие рудные тела.

Зависимость абсолютного потенциала не только от тяготеющей массы, но и, как следует из (2), от её размера - более наглядна при решении проблемы, которую можно назвать «Гравитационное поле тел переменной плотности» или «Задача пульсирующего шара»:
Пусть радиус тяготеющего однородного по плотности шара – есть функция времени. Тогда функция абсолютного гравитационного потенциала в любой фиксированной координате $r > R$ в соответствии с (2) будет иметь вид:

$$U(t) = \left(\frac{2}{3}\right)\gamma\rho\pi R^2 - \frac{\gamma M_R}{r} + \frac{\gamma M_R}{R(t)}$$

Умножив и разделив первое слагаемое на $2R$, имеем ясный результат:

$$(4-2-1) \qquad U(R,t) = \frac{\gamma M_R}{2R} + \frac{\gamma M_R}{R} - \frac{\gamma M_R}{r} = -\frac{\gamma M_R}{r} + \left(\frac{3}{2}\right)\frac{\gamma M_R}{R(t)}$$

Большинство геологических объектов имеют неменяющиеся размеры в реальном (негеологическом) времени. Или меняющиеся столь незначительно, что исследуемая $[(dU/dR(t))] \to 0$. Однако есть и перспективные исключения. Прежде всего, это очаги готовящихся землетрясений. Известен и хорошо исследован рост отрицательной гравитационной аномалии в преддверии даже низкомагнитудных землетрясений. Глобальные сейсмособытия, подобные суматранскому 2004 года, показали в зоне очага весьма заметный рост отрицательной гравианомалии за год до катаклизма. В работах [9] и [10] была представлена и просчитана по некоторым параметрам модель газо-полостного очага землетрясений. Было показано, что газо-заполненные полости размером $10^3$ м. могут вспарывать вышележащий геомассив за счет подъемной силы, давление которой достигает $10^7$ Па. Особенностью газополостного очага как раз является его способность интенсивно аккумулировать астеносферные и литосферные флюиды. При этом размер газовой полости должен интенсивно, а в случае слияния [10] двух или нескольких газовых полостей, скачкообразно увеличиваться. В данной связи правомочно предположить, что в эпицентральной зоне должно наблюдаться уменьшение напряженности граviполя в комплексе с ростом отрицательной аномалии абсолютного потенциала и, соответственно, ростом положительной аномалии $\boldsymbol{v}_S$. Корреляция этих гравиметрических аномалий может дать точную геофизическую картину процесса, а возможно и краткосрочный прогноз сейсмособытия.

Однако, есть только один способ не показать, а доказать перспективность применения $\boldsymbol{v}_S$ гравиметрии и абсолютного потенциала в выявлении таких сложных объектов, как очаг землетрясения. Необходимо решить прямую и обратную геофизическую задачу для глобального геологического объекта.

\*\*\*
Решим прямую задачу и найдём максимальный гравитационный потенциал (МГП) Земли. Как следует из (1) и (2) $U_\infty$ есть функция массы и радиуса нашей планеты. Рассмотрим кинематику падающей из бесконечности пробной массы в интервале $[R., 0]$ - внутри геоида. Очевидно, что процесс падения с граничной поверхности к центру поля даёт для кинетического, а значит и гравитационного потенциала параболическую функцию:

$$(4-2-2) \qquad U_i = \frac{v_{R0}{}^2}{2} = \left(\frac{4}{3}\right)\gamma\rho\pi\int_0^R rdr = \left(\frac{2}{3}\right)\gamma\rho\pi r^2$$

Где: $\rho$ - средняя плотность Земли.

Эквивалентность радиального $\mathbf{v}_{\infty r}$ и тангенсального $v_S$ представлений первой космической скорости за пределами масс-источника, на его граничной поверхности, как известно, заканчивается. Внутри тяготеющей массы наблюдается их очевидное расхождение:

$$(4-2-3) \qquad 0 = \lim_{r \to 0} v_S \neq \lim_{r \to 0} \mathbf{v}_{\infty r} = \boldsymbol{max}$$

что приводит к неравенству кинетического потенциала и «эквипотенциала»:

$$(4-2-4) \qquad \frac{\mathbf{v}_{R0}{}^2}{2} > \frac{v_S{}^2}{2} ., \qquad K_{R0} > C_{R0}$$

Данное неравенство, с позиции закона сохранения энергии, определяется включением в потенциально-кинетический баланс внутренней (упругой) энергии тяготеющего тела. Предположив, что из бесконечности притягивается не пробная масса, а граничная поверхность самого масс-источника, например планеты, несложно видеть, что часть энергии гравитационного сжатия такого протопланетного облака обязана расходоваться на увеличение давления и плотности тяготеющего вещества. При сжатии тяготеющая масса как бы выдавливает энергию формируемого гравитационного поля во внешнее пространство - отсюда и зависимость (4-2-1). Поэтому при расчете МГП необходимо учитывать эту величину, которую можно обозначить, как потенциал гравитационного сжатия. Также необходимо учесть, что в неоднородных (с реально наблюдаемым увеличением плотности по глубине) небесных телах гравитационного генезиса потенциал гравитационного сжатия увеличивается не только за счет упругого, но и теплового расширения. Если представить, что гравитационное поле сжатого тела исчезло - освобожденное вещество приобретёт кинетическую энергию - как упругого, так и теплового происхождения. Последнее что нужно учесть: реальные масс-источники поля принципиально отличаются от однородных моделей смещением плотностного максимума внутрь тяготеющего тела на определенную величину $\Delta R$ - что эквивалентно приращению вида (4-2-1). В итоге:

$$(4-3) \qquad (K_{R0} - C_{R0}) = \frac{\sigma_0 V_R}{M_R} + \Delta U_{\Delta R}{}^* + \Delta U_T = \frac{P_G}{\rho_G} = \varphi_G$$

Где: $\Delta U_T$ - добавленный потенциал теплового расширения
$\Delta U_{\Delta R}{}^*$ - добавленный потенциал смещения плотностного максимума.
$K_{R0} = K_\infty - K_R$ - кинетический потенциал поверхности геоида
                   относительно центра масс.
$C_{0R} = v_{1k}{}^2/2$ - эквипотенциал поверхности Земли.
$\sigma_0 V_R$ - упругое напряжение в объеме Земли.
$P_G$ - характерное гравитационное давление.
$\rho_G = M_R/V_R$ - средняя плотность внутри геоида.
$\varphi_G$ - потенциал гравитационного сжатия Земли.

Если проследить кривую внутреннего эквипотенциала $C_{R0}$ несложно видеть, что её априорная монотонность предписывает (в процессе расхождения с кривой «внутреннего» кинетического потенциала $K_{R0}$) точку максимума с нулевой производной согласно

(невеликой) теореме Ферма. Эту особую точку обозначим $R_0$.

Вернёмся к прямой задаче и пробной массе, которая, оказавшись, в центре геоида, теряет гравитационный потенциал, но приобретёт максимальный кинетический потенциал:

$$(4-4) \qquad \lim_{r \to 0} K_i = \lim_{r \to \infty} U_i$$

Достаточный, чтобы вновь уйти в бесконечность и обратится там в ноль, так что в процессе всего движения из бесконечности в центр масс Земли и обратно, выполняется простое (удовлетворяющее тяготеющей точке) условие:

$$(4-5) \qquad U_i + K_i = U_\infty = K_0 = const$$

Для вычисления МГП Земли, воспользуемся (2), (4-3) и (4-5) приняв $r = R$:

$$(5-0) \qquad U_\infty = U_R + K_R = U_R + C_R + \varphi_G = \left(\frac{2}{3}\right)\gamma\rho\pi R^2 + \frac{v_{1к}^2}{2} + \frac{P_G}{\rho_G}$$

Вычисления по данным 2000 года Университета Вашингтона в Сиэтле о средней плотности и гравитационной постоянной, а также по табличным данным о среднем радиусе и первой космической скорости на поверхности геоида - показывают максимального значения абсолютного гравитационного потенциала Земли:

$(5-1)$ $U_R = 3{,}1843 \cdot 10^7$ Дж/кг., $C_R = 3{,}1284 \cdot 10^7$ Дж/кг., $\varphi_G = 4{,}8491 \cdot 10^7$ Дж/кг

$$(5-2) \qquad U_\infty = 11{,}1652 \cdot 10^7 \text{ Дж/кг}$$

Давление гравитационного сжатия ($P_G = 2{,}7230 \cdot 10^{11}$ Па) определялось в точке перегиба ($\text{grad}P = max$) известной кривой распределения давления внутри геоида. Основное уравнения гидродинамики (В.И.Смирнов. Курс высшей математики):

$$\frac{df}{dV} - \text{grad}P = \rho w$$

показывает, что $(df/dV)$ - объемная плотность массовых сил в веществе геосреды под действием гравитационного сжатия должна достигнуть равновесного изостатического максимума. Ускорение $w$ геосреды - в ситуации изостатического баланса - равно нулю. Отсюда:

$$(5-3) \qquad \frac{df_{\max}}{dV} = \text{grad}_{max}P$$

Достоверность решения прямой геофизической задачи и найденного МГП Земли можно подтвердить, решив на его основании обратную геофизическую задачу. То есть, выявить, при её решении, некий глобальный геологический слой (или границу двух геослоёв, а возможно и геосфер) в котором потенциальная функция имеет характерное значение: достигает экстремума, проходит точку перегиба, пересекает нулевое значение). При этом, наилучшим подтверждением будет геологическая граница или слой выявленный негравитационными методами.

\*\*\*

Решим обратную геофизическую задачу.

Проинтегрируем (2) при условии, что масс-источник сжат в точку ($R = 0$):

$$\gamma \int_{0_\infty}^{R} \frac{M_r}{r^2}\, dr = 0$$

$$(6-1) \qquad U(M, r) = \gamma M_0 \int_0^\infty \left(\frac{1}{r^2}\right) dr = -\frac{\gamma M_0}{r} + A(\gamma M_0)$$

Где $A$ - постоянная интегрирования, поскольку интеграл при таких пределах переходит в неопределенный., зависимость $A$ от $\gamma M_0$ также очевидна.

В расчетных формулах для точечных масс-источников постоянную $A$ принимают равной нулю - впрочем при условии, что в дальнейшем оперируют с разностью потенциалов или потенциалом относительно некоторой поверхности уровня, например поверхности Земли. Однако, с абсолютным потенциалом такое упрощение неправомочно, ибо противоречит закону сохранения энергии. Достаточно представить, что при $A = 0$, максимальный гравитационный потенциал (МГП) Земли и электрона были бы эквивалентны, равняясь нулю. Очевидно также, что постоянная $A$ - есть предел потенциальной функции $U(r)$, и отличает МГП полей тяготения различных масс.
Учитывая симметрию пределов, получаем ясно выраженную закономерность:

$$(6-2) \qquad \lim_{r \to \infty} U(r) = \lim_{r \to 0} K(r)$$

Искомым пределом потенциальной функции таким образом, можно полагать максимальное значение кинетического потенциала:

$$(6-3) \qquad \lim_{r \to R_0} K(r) = \frac{\gamma M_0}{R_0}$$

Где: $R_0$ - координата, в которой $U(M_0 R_0) = 0$ потенциальная функция точечного масс-источника переходит из положительной области в отрицательную.

Таким образом, мы получаем второе указание на наличие особой точки $R_0$.
Обозначив $R_0 -$ «характерный гравитационный радиус», определим его и для неточечного масс-источника, где чисто математическая абстракция с отрицательным абсолютным потенциалом должна перейти в физическую реальность, основанную на законе сохранения энергии. Будем считать, что выявление $R_0$ Земли по МГП и есть заявленная обратная задача.
Понятно, что с учетом (6-3), (2) трансформируется соответствующим образом:

$$(6-4) \qquad U(M, r, R_0) = \left[\left(\frac{2}{3}\right)\gamma \rho \pi R^2 - \frac{\gamma M_R}{r} + \frac{\gamma M_R}{R_0}\right]$$

Точка перегиба такой функции - есть $R_0$ причём, выполняется трехвариантная зависимость распределения плотности от соотношения $R$ и $R_0$:

$$(6-5) \quad 1.\left[\frac{d\rho}{dR} = 0, \ R_0 = R\right].,\ 2.\left[\frac{d\rho}{dR} < 0, \ R_0 < R\right].,\quad 3.\left[\frac{d\rho}{dR} > 0 \quad R_0 > R\right]$$

Для Земли, очевидно, выполняется второе неравенство. Примем во внимание, что в $R_0$ потенциальная функция Земли переходит из положительной параболы в отрицательную гиперболу, при этом её градиент (напряженность поля) достигает максимума, вторая производная (дивергенция напряженности) равна нулю. Не требует доказательств, что приуроченная к $R_0$, столь ярко выраженная эквипотенциальная поверхность ($S_0 = 4\pi R_0^2$) должна коррелироваться с данными глубинного сейсмозондирования. Проведем вычисление:

$$(7-0) \qquad R_0 = \frac{\gamma M_R}{U_\infty} = 3{,}5710 \cdot 10^6 \text{м}.$$

Феномен однородности Земли (совпадение ускорения свободного падения и первой космической скорости полученных из расчета однородной Земли с реальными данными) подсказывал если не эквивалентность, то близость значений $R$ и $R_0$. Вариантом могла бы быть также близость $R_0$ к радиусу внутреннего ядра, где по принятым представлениям плотность максимальна. Результат (7-0) показывает точку перегиба функции абсолютного потенциала на глубине 3571 км. – что соответствует слою D''

Таким образом, граница мантии и ядра Земли, где по многочисленным и выверенным данным наблюдается скачок плотности и фазового состояния вещества, скачок скорости продольных и падение «в ноль» скорости поперечных сейсмических волн, где по последним данным ГСЗ выявлен маломощный, но четко выраженный пограничный слой - является также и ярко выраженной гравитационной границей. Здесь абсолютный гравитационный потенциал имеет точку перегиба функции радиального распределения, напряженность гравитационного поля достигает максимума, а дивергенция вектора напряженности (вторая производная потенциальной функции) равна нулю.

Данное, «вскрывшееся» посредством чисто теоретического анализа, обстоятельство безусловно должно быть всесторонне проверено практикой. Однако уже сейчас ясно, что (7-0) дает такое радиальное распределение гравитационного поля в теле Земли, при котором слой D'' является гравитационно-плотностным максимумом весьма небольшой мощности - порядка $2 \div 3 \cdot 10^5$ м.

При такой мощности слоя D'' кривая распределения напряженности поля в геоиде интенсивно падает до среднего значения 1,16 м/с², которое можно определить из уравнения равновесия поверхности внутреннего или малого ядра Земли:

$$\mathbf{g}_{\text{ср}} = P_{\text{мя}} S_{\text{мя}} / (M_З - M_{\text{мя}}) = 1{,}160 \text{ м/с}^2$$

Где: $(P_{\text{мя}} S_{\text{мя}})$ - сила давления на поверхность малого ядра., $(M_З - M_{\text{мя}})$ - масса Земли за вычетом малого ядра.

Весьма важно, что подынтегральная площадь кривой напряжённости поля в теле Земли, определяющая абсолютный гравитационный потенциал внутри геоида, не превышает абсолютный потенциал на поверхности «однородного эквивалента Земли». Таким образом, выполняется базовый принцип теории поля, которым мы и завершим:

Абсолютный гравитационный потенциал любой элементарной массы, составляющей тяготеющее тело и движущейся лишь под действием его гравитации, стремится к минимуму - вследствие чего гравитационный потенциал граничной поверхности тяготеющего тела при любом плотностном распределении внутри него, не может превышать абсолютный потенциал поверхности однородного по плотности эквивалента:

$$(7-1) \qquad \gamma \int_0^R \frac{M_r}{r^2}\, dr \;\leq\; \left(\frac{2}{3}\right) \gamma \rho_0 \pi R^2$$